# Observation of the acceleration of light in a tapered optical fiber

**Hui Ge, Chong Sheng, Shining Zhu, and Hui Liu***

*National Laboratory of Solid State Microstructures, School of Physics, Collaborative Innovation Center of Advanced Microstructures, Nanjing University, Nanjing 210093, China*

* liuhui@nju.edu.cn

**Abstract:** One of the most fascinating aspects of quantum fields in curved spacetime is the Unruh effect. The direct experimental detection of Unruh temperature has remained an elusive challenge up to now. Gradient optical waveguides manipulating the dispersion of photons are assumed to realize the great acceleration of effective particles, leading to a high effective Unruh temperature. However, experimentally achieving this optical waveguide has not yet been reported. In this work, we exploit a tapered fiber to simulate the accelerated motion of effective particles and obtain an effective Unruh temperature. When light propagating in a tapered fiber is affected by the external high refractive index medium, a leaky phenomenon akin to bremsstrahlung will be observed, and the pattern of leaky radiation is dependent on the acceleration of photons. During the experiments, different accelerations corresponding to different Unruh temperatures are achieved by controlling the shape of the tapered waveguide.

## 1. Introduction

The Unruh effect [1,2] (also known as the Fulling-Davies-Unruh effect) is a prediction that an accelerating observer will treats its surrounding environment as a thermal radiation bath, which is similar to Hawking radiation, and the Unruh temperature is proportional to the observer's acceleration. The Unruh effect has several important applications, for example verifying the excitation of accelerated detectors and atoms, weak decay of non-inertial protons, and the bremsstrahlung of accelerated particles [3]. Due to the high acceleration required for the Unruh temperature to reach a measurable value, it is difficult to directly detect the Unruh effect with the current experimental technology. Recently, an experimental analog of the Unruh effect has been studied in various systems, such as water waves [4], Bose-Einstein condensates [5,6], cold atoms [7,8] and electrons [9], etc. In addition, simulating the gravitational field in the optical system to study some astronomical phenomena that cannot be directly observed is also gradually being developed [10-16], such as simulations of photonic black holes [17-20], Einstein rings [21], and cosmic strings [22], etc. Some scientists have also simulated Hawking-

Unruh radiation in optical systems [23,24]. Here, we design a tapered waveguide with gradient refractive index by means of transformation optics to simulate the Unruh effect.

The Unruh temperature perceived by an accelerating observer can be expressed by the formula

$$T_u = \frac{\hbar a}{2\pi k_B c} \tag{1}$$

Where $a$ is the acceleration of the observer, $\hbar$ is the Planck constant, $k_B$ is the Boltzmann constant, and $c$ is the speed of light in vacuum. When the Unruh temperature reaches room temperature, an acceleration of about $10^{23} m/s^2$ is required. However, a particle acceleration of that magnitude is very difficult to achieve in real experiments. There is a study which exhibits that photons in the waveguide behave as massive quasi-particles [25], and that effective particle acceleration satisfies the formula

$$a = -c^2 \frac{dn}{dz} \tag{2}$$

Where $n$ is effective index in the optical waveguide and $z$ is propagation direction. Smolyaninov et al. further proposed that a high Unruh temperature of up to 30000K can be obtained in a tapered hyperbolic metamaterial waveguide [26], but such a waveguide has not yet been fabricated in practice. Moreover, there is currently no optical measurement method to detect the effective acceleration in the waveguide. In this study, the leaky radiation of tapered fiber is exploited to measure the photons acceleration. Since the speed of light varies with the effective refractive index, photons propagating in the tapered fiber can be used as a simulation for the accelerated motion of particles. We can use this leaky radiation phenomenon to indirectly obtain a very low Unruh temperature.

## 2. Theorical basis

To construct the light acceleration in the optical waveguide, we study the relationship between the effective index and the size of the tapered fiber. The tapered fiber is fabricated by heating and drawing a uniform optical fiber over hydrogen flame, and is characterized under the focused ion beam as shown in Fig. 1(a). It can be seen that the radius of the tapered fiber becomes evenly thinner from left to right. The $z$-coordinate is along the symmetry axis of the tapered fiber, and the relationship between radius $r$ and $z$ is represented by

$$\frac{\Delta r}{\Delta z} = 1.64 \times 10^{-3} \tag{3}$$

Since the thickness of the tip of the tapered fiber is just a few hundred nanometers, only one electromagnetic wave mode transmits inside it. As the radius of the tapered fiber decreases, the corresponding effective refractive index also decreases. The effective refractive index of the tapered fiber is defined as $n_1$, and the relationship between $n_1$ and $z$ is shown in Fig. 1(b). The propagation velocity of a photon in a waveguide is determined by the effective refractive index, which is represented by the formula

$$v = \frac{c}{n_1} \tag{4}$$

Fig. 1(c) illustrates the variation of the group velocity of photons in a tapered fiber with $z$. Therefore, photons propagate at an accelerated speed in the tapered fiber. Finally, the acceleration at different positions is shown in Fig. 1(d).

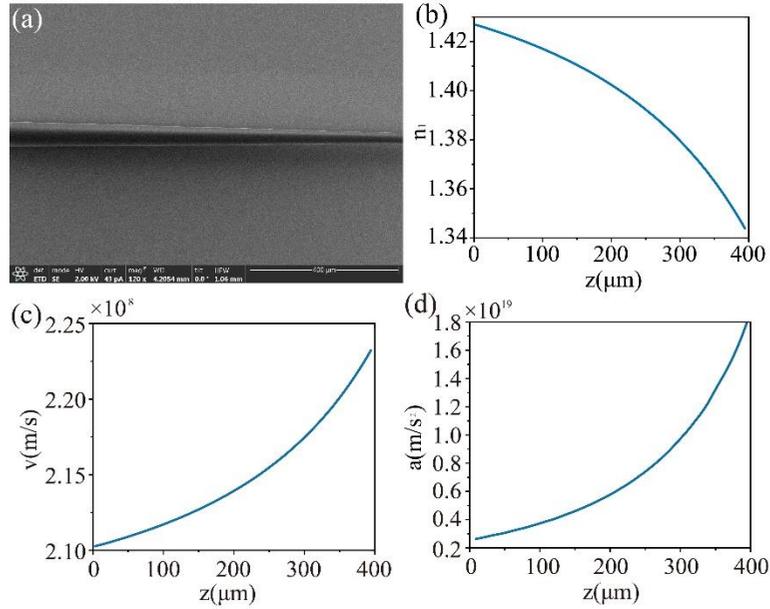

FIG. 1. (a) FIB picture of a tapered fiber. (b) The effective refractive index of the tapered fiber varies with $z$-coordinate. (c) The group velocity of photons in the tapered fiber varies with $z$-coordinate. (d) The acceleration of simulated particles in the tapered fiber varies with $z$-coordinate.

In order to observe the accelerated radiation of light, we use the method of leaky radiation from optical microfibers. As a light transmission tool, optical microfiber has been applied to many aspects such as optical sensors [27-29], optical filters [30], interferometers [31], etc. In general, the effective refractive index of optical microfiber should be greater than that of its surrounding medium, thereby the light beam is confined to propagate in the optical microfiber. However, when the effective refractive index of the optical microfiber is smaller than that of its surrounding medium, it causes light leaking out. Assuming the effective refractive index of the microfiber and the surrounding medium are $n_1$ and $n_2$ respectively, if $n_2 > n_1$, then the light travelling through the microfiber will leak out. According to Snell's refraction law, the radiation angle $\theta$ can be calculated with the formula

$$\cos\theta = \frac{n_1}{n_2} \tag{5}$$

Due to the fact that an optical microfiber with uniform thickness has a constant refractive index, the refractive index of the environmental medium can be detected by measuring the angle of the leaky radiation [32]. Conversely, if the refractive index of the environmental medium is known, then the effective refractive index of microfibers with different thicknesses can also be detected. Therefore, this method can be used to calculate the effective refractive index at different positions on the tapered fiber.

## 3. Experimental structure and phenomena

To observe the radiation phenomenon, we design a slab waveguide with a high refractive index. The slab waveguide is composed of a quartz substrate, a polymethyl methacrylate (PMMA) film and a silver film. First, we spin-coat PMMA polymer on a silica ($SiO_2$) substrate. An appropriate amount of the rare earth ions are mixed uniformly into the PMMA polymer so that fluorescence imaging can be carried out. We choose $Eu^{3+}$ as it absorbs light in the purple wave band, and re-emits fluorescent light in the red wave band. This means that the radiated light propagation in the waveguide can be observed. After drying the spin-coated sample in an oven, we then measure the thickness and refractive index of the PMMA layer with an ellipsometer. The measured thickness of the PMMA film is 1081nm, and the refractive index of the PMMA film at a wavelength of 405nm is 1.62713. Since the substrate and PMMA layer are transparent materials, we finally deposit a 30nm thick Ag film over the PMMA in order to observe a clear radiation phenomenon. When considering the PMMA layer thickness and refractive index in the eigenmode equation of the slab waveguide, we are able to obtain the effective refractive index of the PMMA waveguide as $n_2 = 1.618$. Clearly, the effective refractive index of the PMMA waveguide is larger than that of the optical fiber.

We design the experimental structure shown in Fig. 2(a) to excite the leaky radiation of a uniform microfiber. The microfiber is attached to the PMMA waveguide surface, and a 405nm laser is coupled into the microfiber via a fiber collimator. Since the laser beam propagating in the microfiber is directly affected by the PMMA waveguide, it radiates out of the microfiber and propagates inside the PMMA waveguide. We place the sample on the platform of a microscope equipped with an sCMOS (ProgRes CFscan), the radiation transmitted within the PMMA waveguide can be observed through a red filter. A clear image of the radiation captured by sCMOS is shown in Fig. 2(c), the radiation angle at different locations has the same value. The measured radiation angle is $21.2°$, and the effective refractive index of the uniform microfiber in the experiment is calculated as 1.508. Following this, we replace the uniform microfiber in Fig. 2(a) with a tapered fiber to obtain the experimental structure shown in Fig. 2(b), and the radiation angle increases as the radius of the tapered fiber decreases. The radiation phenomenon exhibited by the tapered fiber captured in the experiment is shown in Fig. 2(d), which shows that the radiation angle gradually increases from $29.40°$ to $37.38°$. The radiation phenomenon whereby the angles gradually change is similar to bremsstrahlung. By measuring these radiation angles, the effective refractive index gradient can be calculated to obtain the acceleration of light.

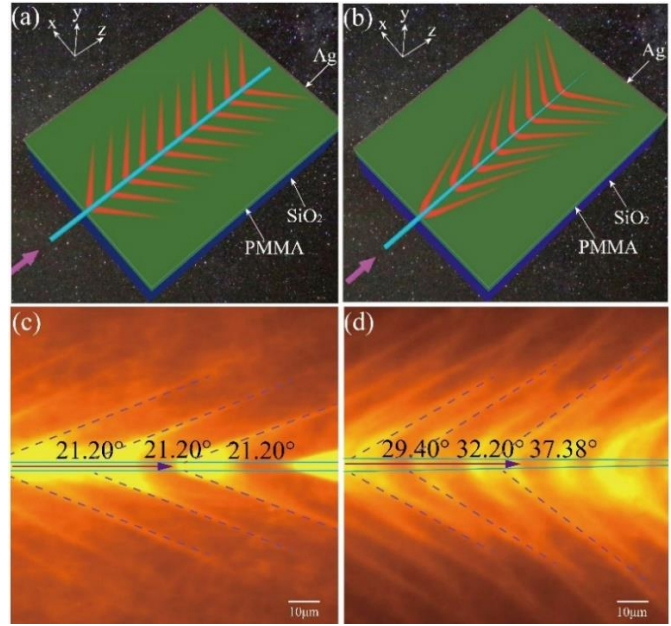

Fig. 2. (a) The experimental structure for exciting leaky radiation of uniform microfiber. The blue part of the slab waveguide is a 1cm×1cm, 1mm thick quartz substrate, the green part represents PMMA film with a thickness of 1081nm, and the gray part represents Ag film with a thickness of 30nm. (b) The experimental structure for exciting leaky radiation of tapered fiber. (c) The radiation phenomenon generated by a uniform microfiber. The radiation angles at different locations remain the same value. (d) The radiation phenomenon generated by a tapered fiber. The waveguide tapers evenly from left to right, and the radiation angles increase accordingly.

## 4. Experimental results and discussion

As shown in Fig. 3(a)-(d), the leaky radiation experiments are carried out with tapered fibers of different shapes. In this study, we find that the more dramatically the shape of the tapered fiber, the more obvious change in the radiation angle along the different locations of the tapered fiber is. By referring to the size of the tapered fiber, we calculate the variation in the radiation angle excited by each tapered fiber with $z$, and draw it with the radiation angle measured during the experiments on the same diagram. As shown in Fig. 3(e)-(h), the experimental value matches the theoretical value.

To obtain Unruh temperature which is simulated by the tapered microfiber, we take Fig. 3(b) for instance, which undergoes a radius reduced from $1.31 \mu m$ to $0.93 \mu m$, within the range of $z$ change of $232 \mu m$. We measure the radiation angle at different locations in the picture to achieve the effective refractive index at the corresponding position of the tapered fiber according to Eq. (5). How to accurately measure the radiation angle is described in the supplementary material. The calculated refractive index difference is $\Delta n_1 = 0.44$, the particle acceleration simulated by the tapered fiber in Fig. 3(a) is equal to

$$a = -c^2 \frac{dn_1}{dz} = -c^2 \frac{\Delta n_1}{\Delta z} = 5.63 \times 10^{19} m/s^2 \qquad (6)$$

which means the Unruh temperature is $T_u = 0.230K$.

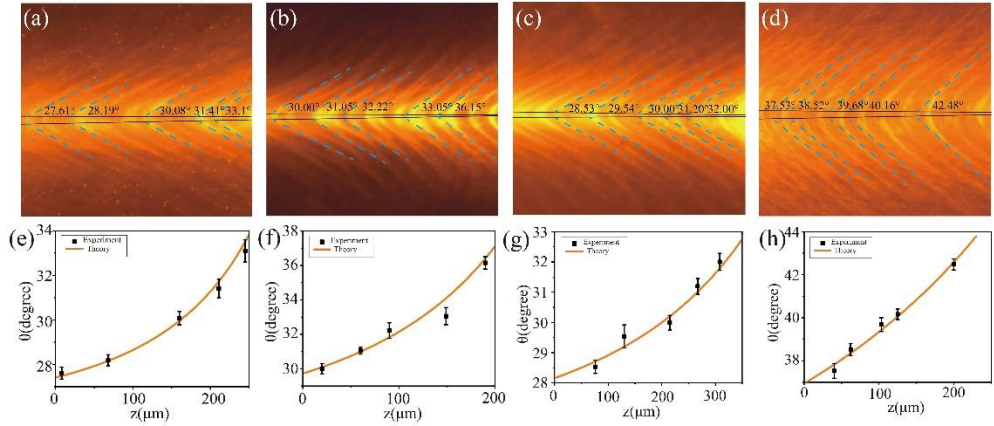

Fig. 3. Experimental phenomenon and data processing (a)-(d) Radiation phenomenon of tapered fibers with different shapes observed at microscope. (e)-(h) Multiple sets of data are taken at different locations on tapered fibers, the experimental data and the theoretical data are drawn on the same diagram for comparison. The black square represents experimental radiation angle and the orange line represents theoretical radiation angle.

We use the gradient of the radius on the $z$-coordinate to represent the shape of different tapered fibers and define it as

$$\eta = \frac{\Delta r}{\Delta z} \qquad (7)$$

The effective particles acceleration and the corresponding Unruh temperature simulated by different tapered fibers are calculated and are listed in Table1. The larger $\eta$ is, the higher the simulated Unruh temperature.

**Table1: Unruh temperature simulated by tapered fibers of different shapes**

| $\eta$ | $3.48 \times 10^{-3}$ | $2.24 \times 10^{-3}$ | $1.64 \times 10^{-3}$ | $1.00 \times 10^{-3}$ |
|---|---|---|---|---|
| $a(m/s^2)$ | $5.63 \times 10^{19}$ | $4.02 \times 10^{19}$ | $2.65 \times 10^{19}$ | $1.62 \times 10^{19}$ |
| $T_u(K)$ | 0.230 | 0.164 | 0.108 | 0.066 |

According to the relationship between the Unruh temperature and effective refractive index gradient, and the relationship between effective refractive index and radius, the theoretical Unruh temperature is calculated as shown in Fig. 4, the experimental data agrees with it. We obtain different Unruh temperatures by preparing different tapered fiber shapes and analyzing the variation in radiation angles in the radiation patterns. Therefore, through our experimental method, different Unruh temperatures can be obtained by controlling the shape of the tapered fiber.

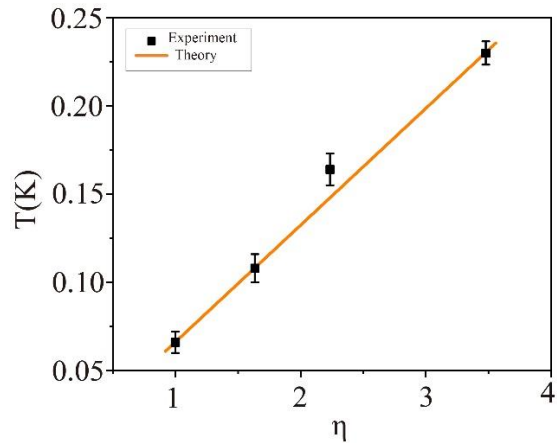

Fig. 4. Simulated Unruh temperature varies with the shape of the tapered fiber. The black square represents experimental data and the orange line represents theoretical data.

## 5. Conclusion

In this work, we obtained the leaky radiation of light by coupling the mode of the tapered microfiber with that of the PMMA waveguide, thereby we realized the observation of photons acceleration. Furthermore, we calculated the Unruh temperature simulated in tapered waveguide, and obtained the relationship between the Unruh temperature and the shape of the tapered fiber, which will be helpful for future research on the experimental observation of the Unruh effect in optical waveguides. Our experimental method can control the accelerated propagation of photons in the transformation optical waveguide, and provides a new solution for regulating the group velocity of optical signals in the integrated photonic chip.

**Acknowledgments.** This work was financially supported by the National Key R&D Program of China (Grants No. 2017YFA0303702 and No. 2017YFA0205700), the National Natural Science Foundation of China (Grants No. 11690033, No. 61425018, No. 11621091, and No. 11704181), and the Fundamental Research Fund for the Central Universities, China (Grant No. 14380139).

**Disclosures.** The authors declare there are no conflicts of interest related to this article.

**Data availability.** Data underlying the results presented in this paper are not publicly available at this time but may be obtained from the authors upon reasonable request.

**Supplemental document.** See Supplement 1 for supporting content.